\documentclass[twocolumn,showpacs,superscriptaddress,amsmath,amssymb,prl]{revtex4}

\usepackage{graphicx}%
\usepackage{dcolumn}%

\begin{document}

\title{Non-monotonic pseudo-gap in high-$T_c$ cuprates}

\author{A. A. Kordyuk}
\affiliation{IFW Dresden, P.O. Box 270116, D-01171 Dresden, Germany}
\affiliation{Institute of Metal Physics of National Academy of Sciences of Ukraine, 03142 Kyiv, Ukraine}

\author{S. V. Borisenko}
\author{V. B. Zabolotnyy}
\author{R. Schuster}
\author{D. S. Inosov}
\affiliation{IFW Dresden, P.O. Box 270116, D-01171 Dresden, Germany}

\author{\\R. Follath}
\author{A. Varykhalov}
\affiliation{BESSY GmbH, Albert-Einstein-Strasse 15, 12489 Berlin,
Germany}

\author{L. Patthey}
\affiliation{Swiss Light Source, Paul Scherrer Institut, CH-5234
Villigen, Switzerland}

\author{H. Berger}
\affiliation{Institut de Physique de la Mati\'ere Complexe, EPFL, 1015 Lausanne, Switzerland}

\date{Nov 1, 2007}%

\begin{abstract}
The mechanism of high temperature superconductivity is not resolved for so long because the normal state of
cuprates is not yet understood. Here we show that the normal state pseudo-gap exhibits an unexpected non-monotonic
temperature dependence, which rules out the possibility to describe it by a single mechanism such as
superconducting phase fluctuations. Moreover, this behaviour, being remarkably similar to the behaviour of the
charge ordering gap in the transition-metal dichalcogenides, completes the correspondence between these two
classes of compounds: the cuprates in the PG state and the dichalcogenides in the incommensurate charge ordering
state reveal virtually identical spectra of one-particle excitations as function of energy, momentum and
temperature. These results suggest that the normal state pseudo-gap, which was considered to be very peculiar to
cuprates, seems to be a general complex phenomenon for 2D metals. This may not only help to clarify the normal
state electronic structure of 2D metals but also provide new insight into electronic properties of 2D solids where
the metal-insulator and metal-superconductor transitions are considered on similar basis as instabilities of
particle-hole and particle-particle interaction, respectively.
\end{abstract}


\preprint{\textit{xxx}}

\maketitle

\begin{figure*}[t]
\includegraphics[width=14cm]{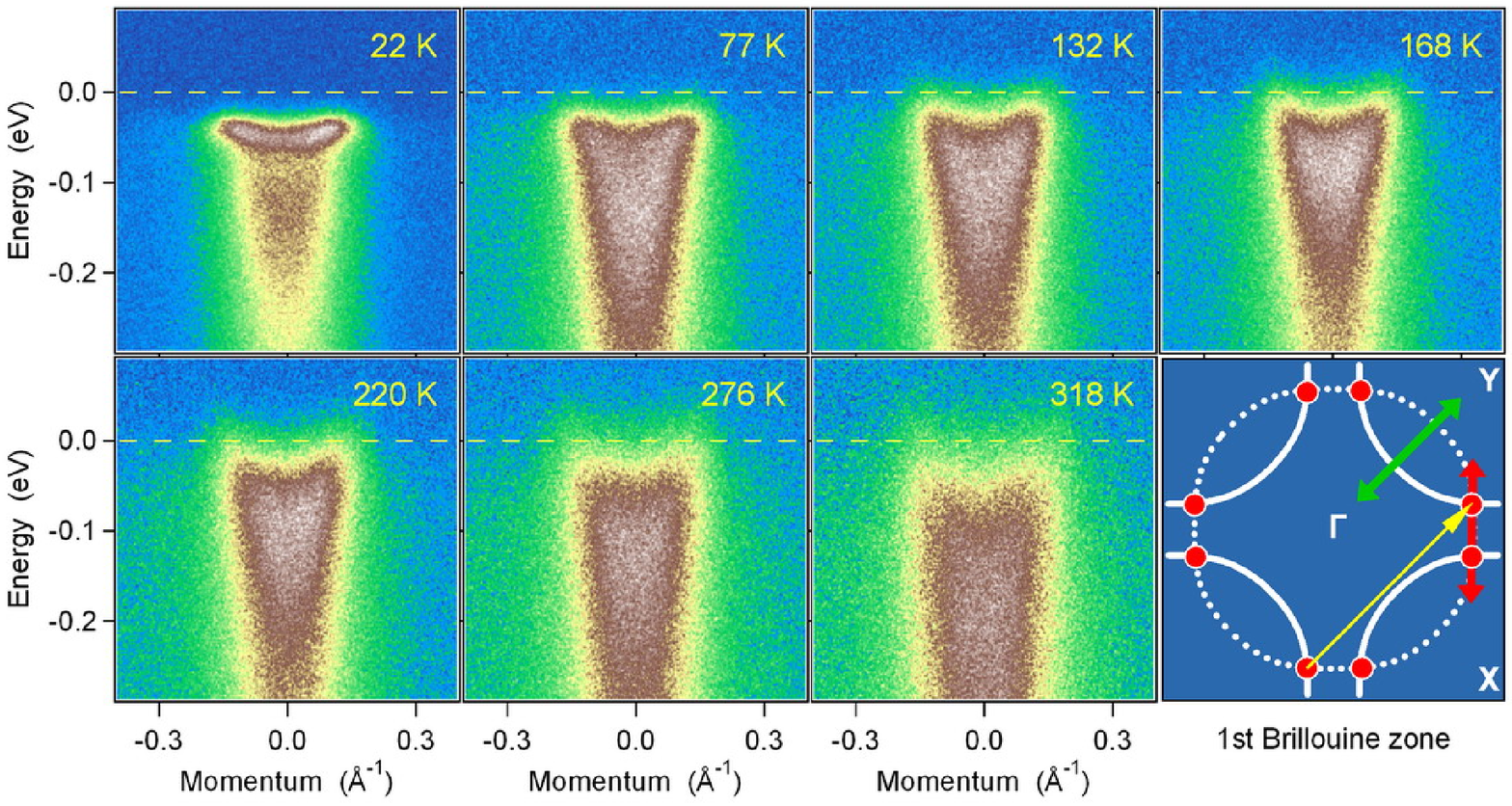}
\caption{\label{Fig1} \textbf{Temperature evolution of the ARPES spectra.} Panels in terrain colorscale, which
represent the "ARPES images" (the momentum distribution of the photoemission intensity) taken along the
"hot-spots" cross-section (see the last panel), demonstrate the evolution of the quasiparticle spectral weight in
a wide temperature range from 20 K to 320 K. All spectra are measured at 50 eV excitation energy at which only the
antibonding band is visible. The sketch of the Fermi surface (FS, white solid lines) illustrates the position of
the "hot spots" (red points), which are the points on the FS connected by the antiferromagnetic ($\pi, \pi$)
vector (yellow arrow); the dotted line represents the FS shifted by the ($\pi, \pi$) vector; the red and green
double headed arrows show the position of the hot-spots crossing and nodal crossing, respectively.\\ \\ }
\includegraphics[width=15cm]{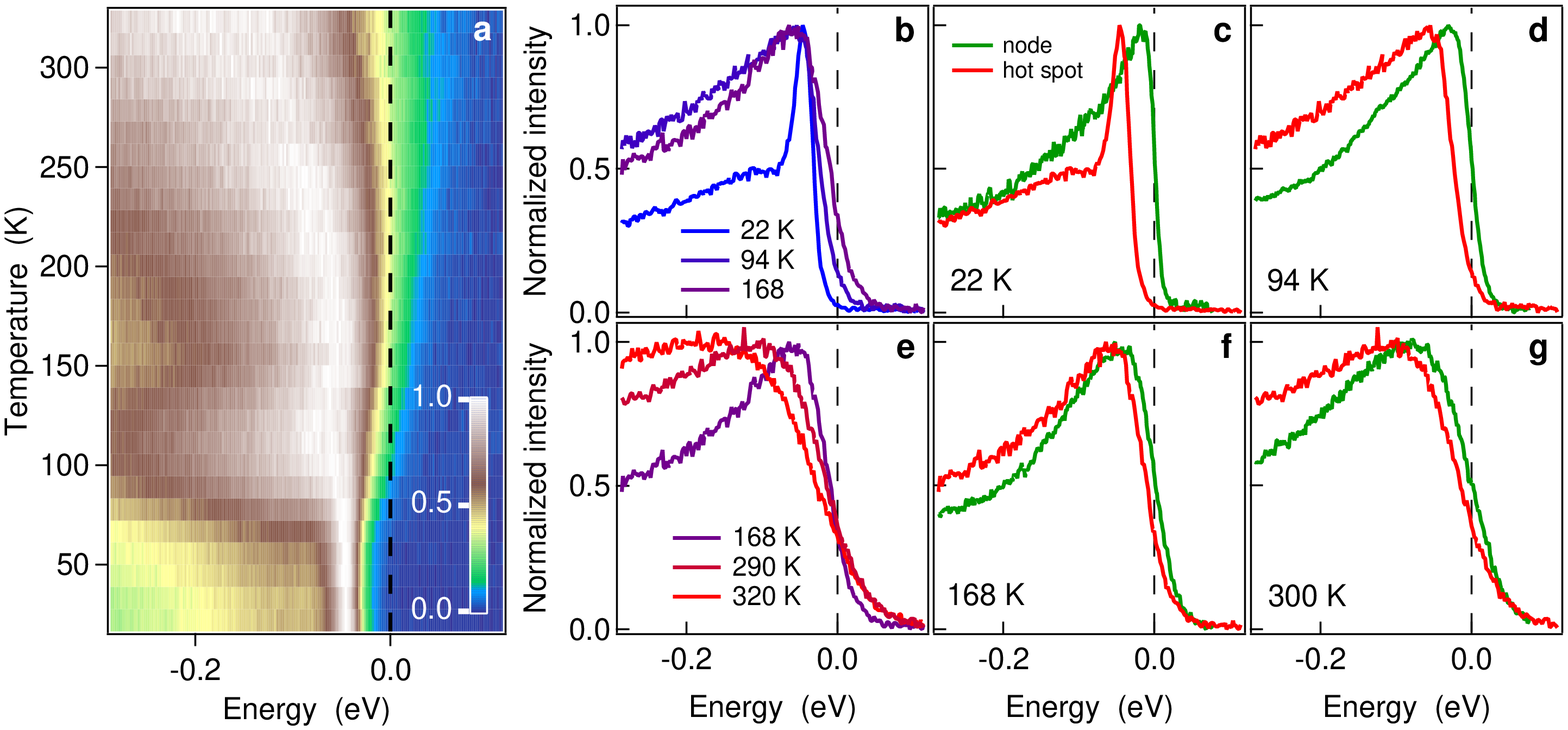}
\caption{\label{Fig2} \textbf{The temperature map.} (a) The temperature map which consists of a number of momentum
integrated energy distribution curves (EDCs) measured at different temperatures at a "hot spot". Separate EDCs are
shown in panels (b-g): as compared to each other (panels b and e) and to the similar EDCs measured each for the
same temperature but along the nodal direction (panels c, d, f, g). The gap is seen as a shift of the leading edge
midpoint (LEM). In terms of the colorscale of panel a, the LEM corresponds to the boundary between yellow and
brown close to the Fermi level. All the "hot spot" EDCs are integrated in momentum range $\pm$ 0.15 {\AA}$^{-1}$
around $k_F$.}
\end{figure*}

\begin{figure*}[t]
\includegraphics[width=11cm]{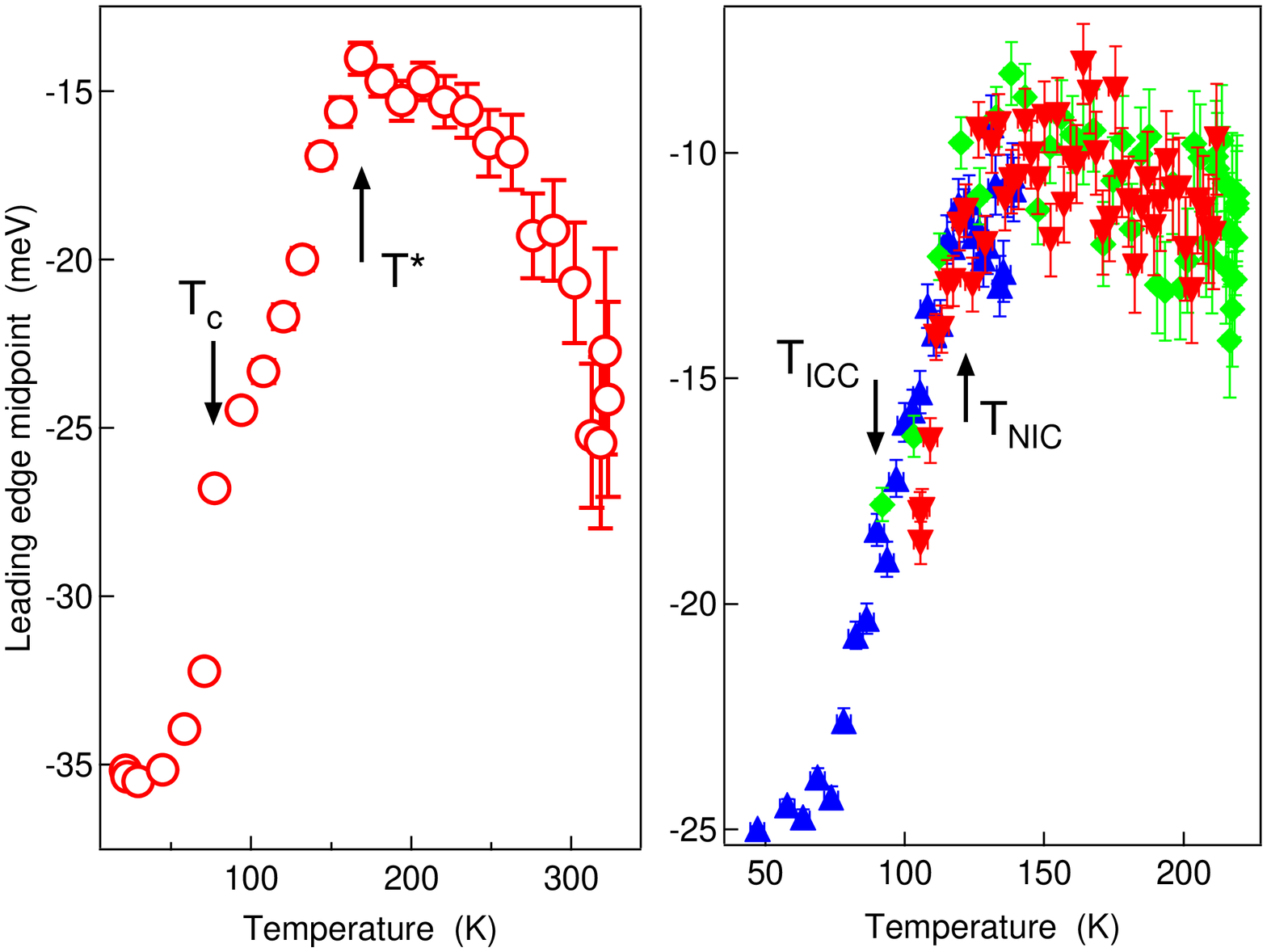}
\caption{\label{Fig3} \textbf{Non-monotonic gap function.} The position of the leading edge midpoint (LEM) of the
integrated $k_F$ EDCs (averaged for two Fermi-crossings), as function of temperature for an underdoped Tb-BSCCO
(left) with $Tc$ = 77 K and $T^*$ = 170 K is remarkably similar to the pseudo-gap in a transition-metal
dichalcogenide TaSe$_2$ (right) \cite{6} with the transitions to the commensurate and incommensurate CDW phases at
$T_{ICC}$ = 90 K and $T_{NIC}$ = 122 K, respectively.}
\end{figure*}

In the normal state, the conventional superconductors are metals and the superconductivity emerges as a
particle-particle instability, known as Cooper pairing, resulting in a superconducting gap (SG) in the spectrum of
one-particle excitations. The superconducting cuprates were found to be different, exhibiting a pseudo-gap (PG)
already in the normal state \cite{1}, and, since its discovery, the PG towers as an impregnable wall on the way to
understanding the high-$T_c$ superconductivity \cite{2}, either foreshadowing the SG \cite{3} or interfering with
it \cite{4}. In spite of recent progress \cite{5,6,7,8,9} in disentangling the SG and PG phenomena, the origin of
the PG remains unclear. It is still widely believed that the PG is peculiar to the cuprates \cite{2,10} and that
it closes above $T^*$ \cite{10,11} but a characteristic behaviour which can help to uniquely identify its origin
among many possibilities is missing \cite{2,9}. Here we show that the PG in cuprates exhibits an unexpected
non-monotonic temperature dependence, which rules out the possibility to describe it by a single mechanism such as
superconducting phase fluctuations. Moreover, this behaviour, being remarkably similar to the behaviour of the
charge ordering gap in the transition-metal dichalcogenides \cite{12}, completes the correspondence between these
two classes of compounds: the cuprates in the PG state and the dichalcogenides in the incommensurate charge
ordering state reveal virtually identical spectra of one-particle excitations as function of energy, momentum and
temperature. This allows one not only to conclude that the governing contribution to the PG in cuprates comes from
an incommensurate charge ordering but also to consider the PG as a general phenomenon for 2D metals.

Although one of the first theories of a pseudogap \cite{13} was developed for one dimensional charge density wave
(CDW) systems \cite{14}, the PG phenomenon has started to play the lead only after its observation in HTSC
\cite{1}. Now the PG is considered peculiar to HTSC \cite{10} because it is properly anisotropic, resembling the
unique $d$-wave symmetry of the SG, and appears in a solid region in the universal doping-temperature ($xT$) phase
diagram of the cuprates \cite{2}. The deviation of the PG anisotropy from pure $d$-wave dependence have been
discussed earlier in terms of "Fermi arcs" \cite{15}, and recent momentum resolved experiments \cite{5,6,8,9} have
shown that the momentum distributions of the SG and PG are actually different, and one can speak about two gaps
coexisting in the superconducting state. This, however, does not reveal the PG origin \cite{9}. To find the true
mechanism among a number of possibilities is, in part, complicated because the PG phase boundaries on $xT$-phase
diagram, especially the crossover temperature $T^*$, are often ill defined \cite{2}.

The data were collected using the synchrotron radiation (BESSY, "one-cubed" ARPES and SLS, SIS beamline) and the
photoelectron analysers SES R4000 and SES 100. All the presented spectra are measured on one underdoped samples
($x$ = 0.11, with $T_c$ = 77~K): Bi(Pb)$_2$Sr$_2$Ca(Tb)Cu$_2$O$_{8+\delta}$ (Tb-BSCCO) but similar results have
been obtained also for other Tb-BSCCO and Dy-BSCCO samples of similar doping level. The key spectra are measured
along the cut of the Brillouin zone (red double headed arrow on the last panel of Fig.1) which goes through the
"hot spots" (shown by red points) for the antibonding band. The energy distribution curves (EDC) where integrated
over a finite momentum range ($\pm$ 0.15 {\AA}$^{-1}$) around $k_F$. Such an integration ensures that the leading
edge midpoint of a non-gaped spectrum stays at the Fermi level \cite{7,16}.

Figs. 1 and 2 present the results of the detailed temperature dependence of the one-particle excitations spectra
measured by angle resolved photoemission (ARPES). The data are measured along the cut (red double headed arrow in
the sketch of the Fermi surface in Fig.~1) of the Brillouin zone (BZ) which goes through the "hot spots" (the
spots at the Fermi surface which are connected by the antiferromagnetic vector, see the sketch) for an underdoped
Tb-BSCCO. Fig.~1 represents several characteristic spectra taken at different temperatures. Fig.~2 contains
representation of the same dataset as a temperature map (panel a) and as momentum integrated energy distribution
curves (EDCs) measured at different temperatures and compared to each other (panels b and e) as well as to the
similar EDCs but measured for each temperature along the nodal direction (panels c, d, f, g). The gap is seen as a
shift of the whole spectrum from the Fermi level (Fig.1) or, more explicitly, as a shift of the leading edge
midpoint (LEM) of a gapped EDC (Fig.2). Since the momentum integrated EDC of the non-gapped spectrum is expected
to stay at zero binding energy for any temperature \cite{7,16}, as it is observed for the nodal EDCs (Fig.2 c, d,
f, g), the finite shift of the LEM is a good empirical measure for a gap of unknown origin. From the temperature
map presented in Fig.~2a one can easily see an unusual temperature evolution of the gap (in terms of the
colorscale, the LEM corresponds to the right boundary between yellow and brown): first it decreases with
increasing temperature up to about 170 K, then it starts to increase again. It is important to stress that the LEM
is not the only quantity which describes the PG phenomenon. The shift of the LEM is also correlated with the width
of the spectra as well as with the position of EDC maxima (see Fig.2a): the closer the LEM is to the Fermi level
the narrower is the spectrum (compare spectra at 132 K, 220 K to 168 K in Fig.~1). In other words, in the normal
state it is at 170 K when the quasiparticles are most coherent.

The left panel of Fig.~3 summarizes the extracted values of LEM for the two "hot spots" of Tb-BSCCO. The presented
temperature dependences allow to highlight the following observations. First, the PG is clearly observed even at
room temperature, which is well above the pseudo-gap temperature ($\sim$ 200 K) expected for the BSCCO sample with
such doping level \cite{11,17}. Second, the gap evolution with temperature is non-monotonic, reaching a finite
minimum value at about 170 K and increasing on both sides---a new phenomenon which has not been observed in
cuprates before. Third, a sharp change of the slope of the dependence at 170 K suggests a real phase transition at
this temperature (which we denote as $T^*$ here). Alternatively, $T^*$ can be defined as the temperature at which
the quasiparticles are most coherent.

The observed $\Delta(T)$ dependence indicates that the previous understanding of the PG properties was, at least,
not complete. A finite gap above $T^*$ can be considered as a precursor of the PG below $T^*$. This questions some
of PG scenarios, such as superconducting phase fluctuations \cite{10,13}, and restricts the others: the origin of
the preformed gap should be consistent with the PG mechanism.

What is remarkable, the observed non-monotonic temperature dependence of the gap is very similar to the analogous
quantity measured for TaSe$_2$ (Fig.~3, right panel). This similarity appears in the following: (1) the
transitions at $T_{NIC}$ and $T^*$ are similarly sharp (the derivative $d\Delta/dT$ has a finite jump); (2) in the
temperature range from Tc to $T^*$ the gap decreases linearly like from $T_{ICC}$ to $T_{NIC}$; (3) the gap does
not close at $T^*$ and $T_{NIC}$; and (4) increases again above these temperatures.

\begin{figure}[t]
\includegraphics[width=8.6cm]{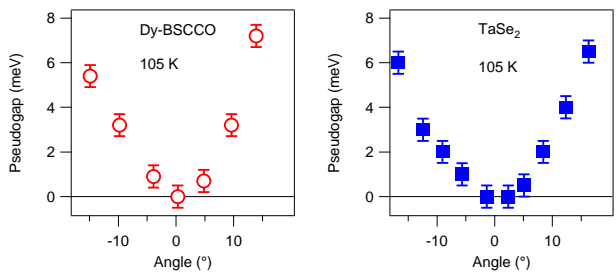}
\caption{\label{FigS1} \textbf{Momentum anisotropy of the pseudo-gap.} The dependences of the gap on the Fermi
surface as function of angle in Dy-BSCCO in the pseudo-gap state (left) and in TaSe$_2$ in the incommensurate CDW
state (right) are almost identical.}
\end{figure}

It is important to stress here that the presented one-to-one correspondence in temperature evolution of the gap in
BSCCO and TaSe$_2$ completes the overall similarity of the spectral functions of these two classes of compounds.
The strong $d$-wave like anisotropy of the PG cannot be considered peculiar to HTSC anymore since similar
dependence have been observed for TaSe$_2$ \cite{12}. Fig.~4 illustrates this. Moreover, the whole spectra of
one-particle excitations for these compounds in the region where the PG opens looks essentially the same: The
depletion of the spectral weight at the Fermi level is really partial and changes smoothly with binding energy;
there is neither bending nor backfolding of the experimental dispersion (the Bogoliubov-type dispersion predicted
by the BCS theory) in the PG state as it is seen in the superconducting state of cuprates \cite{9,18} or in the
commensurate CDW state of TaSe$_2$ \cite{12}. To compare the superconducting and PG ARPES spectra of BSCCO see
Ref.~\onlinecite{18}. This allows one to conclude that the cuprates and the dichalcogenides in the PG state reveal
virtually identical spectra of one-particle excitations as function of energy, momentum and, now, temperature. In
the following, we discuss the analogy between the HTSC and CDW compounds in more details.

It is well known from neutron scattering experiments \cite{20} that TaSe$_2$ at $T_{NIC}$ = 122 K undergoes the
transition into an incommensurate CDW state which turns to the commensurate CDW at $T_{ICC}$ = 90 K. Therefore, in
this temperature region, the observed gap is the incommensurate band gap \cite{21} (IBG). The observed linear
evolution of the IBG may be a consequence of a gradual change of the electronic structure (more precisely, the
nesting vector) with temperature towards the commensurability \cite{12}. At $T_{ICC}$ the CDW becomes commensurate
and the IBG transforms into a well-defined normal band gap. A similar scenario can be valid for the cuprates, with
the only difference that the superconductivity happens before the charge ordering becomes commensurate and the
high-$T_c$ compounds, due to superconducting renormalization of the electronic structure, escape the commensurate
charge ordering. There is one known exception to this trend, the LBCO compound with 1/8 doping level \cite{22},
where the charge ordering suppresses the superconductivity completely \cite{7}.

In order to understand the gap above $T_{NIC}$, in case of TaSe$_2$, or $T^*$, in case of BSCCO, we may suggest
two scenarios: the fluctuating incommensurate charge ordering \cite{13} or a depletion of the spectral weight due
to momentum dependent scattering that can be formulated in terms of a nesting and momentum dependent self-energy
\cite{23,24}. Within the former, one would expect that at higher temperature the fluctuating gap will eventually
close. A "resistivity saturation" observed for underdoped LSCO above 500 K \cite{1,25} may be a signature for it.
In the latter case, the increase of the gap can be related with the temperature evolution of the band structure
towards better (commensurate) nesting \cite{12}, or may be caused by a temperature effect on momentum dependent
scattering which needs to be understood.

To advocate the discussed analogy we note that there is a number of observations of the charge or spin ordering in
the hole-doped cuprates: the 'checkerboard' pattern by tunneling \cite{26} and ARPES \cite{27} experiments,
spatial modulation of spin/charge density by neutron scattering \cite{22}, the charge density modulation by x-ray
scattering \cite{28}. Very recently, the idea of density-wave phase is supported by careful measurements of the
Hall effect \cite{29}.

The project is part of the Forschergruppe FOR538. We acknowledge discussions with J. Fink, B. Buechner, M.
Knupfer, A. Chubukov, I. Eremin, T. Valla, M. Sadovskii, and technical support from R. Huebel.

\end{document}